\documentclass[conference]{IEEEtran}
\IEEEoverridecommandlockouts

\usepackage{cite}
\usepackage{amsmath,amssymb,amsfonts}
\usepackage{algorithmic}
\usepackage{graphicx}
\usepackage{textcomp}
\usepackage{xcolor}
\usepackage{hyperref}
\def\BibTeX{{\rm B\kern-.05em{\sc i\kern-.025em b}\kern-.08em
    T\kern-.1667em\lower.7ex\hbox{E}\kern-.125emX}}

\usepackage{soul}

\def\bF{\mathbf{F}}
\def\bx{\mathbf{x}}

\def\bw{\mathbf{w}}

\begin{document}

\title{Rethinking Chronological Causal Discovery with Signal Processing
\\
\thanks{
This work was supported by the UKRI AI programme and the Engineering and Physical Sciences Research Council, for CHAI - Causality in Healthcare AI Hub [grant number EP/Y028856/1].}
}

\author{\IEEEauthorblockN{Kurt Butler, Damian Machlanski, Panagiotis Dimitrakopoulos and Sotirios A. Tsaftaris}
\IEEEauthorblockA{
School of Engineering, The University of Edinburgh, Edinburgh, UK \\
Causality in Healthcare AI Hub (CHAI), UK \\
\{\texttt{kbutler2; d.machlanski; pdimitra; s.tsaftaris}\}\texttt{@ed.ac.uk}
}
}

\maketitle

\begin{abstract}
Causal discovery problems use a set of observations to deduce causality between variables in the real world, typically to answer questions about biological or physical systems. These observations are often recorded at regular time intervals, determined by a user or a machine, depending on the experiment design. There is generally no guarantee that the timing of these recordings matches the timing of the underlying biological or physical events. In this paper, we examine the sensitivity of causal discovery methods to this potential mismatch. We consider empirical and theoretical evidence to understand how causal discovery performance is impacted by changes of sampling rate and window length. 
We demonstrate that both classical and recent causal discovery methods exhibit sensitivity to these hyperparameters, and we discuss how ideas from signal processing may help us understand these phenomena.
\end{abstract}


\section{Introduction}
Causal information is important because it allows us to reason about the world and the effects of how we interact with the world. In some cases, we may be uncertain of the presence or nature of causal relationships in real systems, and so we wish to infer the existence of causalities through data. One such task is causal discovery, a detection-theoretic task in which one attempts to perform graph topology inference of a causal system \cite{scholkopf2021toward}. Causal graphs aim to represent relationships that generalize when the system is intervened upon, shocked, or otherwise externally manipulated \cite{pearl2009causality}. In this work, our particular interest is in the chronological causal discovery (CCD) task, where the goal is to detect causal influences between signals or time series.

The gold standard for modelling how natural systems evolve in time, and by extension their causal relationships, is differential equations \cite{scholkopf2021toward}. However, instead of relying solely on continuous-time differential equations, we often invoke discrete-time models that abstract the underlying causal relationships \cite{beckers2020approximate}. Many recent works in the causal machine learning (causal ML) literature have focused on discrete-time models, without questioning how faithful these models are to the underlying continuous-time dynamics. If we consider the CCD task to be a problem of understanding the generative process behind a data set, then the mismatch between discrete and continuous becomes clearly important.

In this paper, we revisit the CCD task with a signal processing perspective. We outline a few fundamental concerns about modelling and detecting causal influences with time series, and demonstrate that those concerns apply to both classical and modern causal discovery methods. Finally, we conclude with some reflections on what the signal processing community can offer to the pursuits of causality.

\section{A Primer on Causal Discovery Methods}
\label{sec:methods}
Here we provide a high level review of the chronological causal discovery task, and some representative approaches to the CCD task. Our list is necessarily non-exhaustive, is intended to highlight a representative subset of methods, and we only consider approaches for stationary processes (Methods for non-stationary processes are beyond the scope of the current work).

\paragraph{Granger Causality (GC) and Transfer Entropy (TE)}
While GC is potentially the oldest approach to detecting causal influences \cite{granger1969investigating}, it remains incredibly popular and is still actively discussed \cite{shojaie2022granger}. Although GC has been criticized as not representative of ``true'' causality, due to the possibility of unobserved confounding, the GC principle can be used as a part of ``true'' causal inference under certain assumptions \cite[pp. 202-208]{peters2017elements}. The argument that unobserved confounding biases inference also applies to causal ML algorithms, which is why their assumptions often include some version of causal sufficiency \cite{beckers2021causal}. The classical GC algorithm fits linear autoregressive models to the signals of interest and detects causality when conditioning on the putative cause improves prediction. There are many other variants of GC that change the models being used, or the test statistic used to make a decision.

Transfer entropy (TE), a relative of the GC approach, formulates detection using information theory instead of predictive performance \cite{schreiber2000measuring}. TE measures directed information flow, allowing it to detect nonlinear influences that linear prediction models may miss. TE is argued to be more general than the standard GC approach, since the principles of information theory are innately model agnostic and nonlinear \cite{vicente2011transfer}. For Gaussian systems, TE and GC are even equivalent \cite{barnett2009granger}. Since TE is sometimes thought of as an extension of the GC test, TE depends on the sampling rate and window size in a similar manner to the GC test. 

\paragraph{Constraint-based}
Constraint-based methods are a family of classical approaches to (static) causal discovery that are based on conditional independence (CI) tests to recover a graph skeleton, further refined with a set of constraining rules to orient edges. The Peter-Clark (PC) algorithm is a prime representative of such methods~\cite{spirtes2000causation}. PC has been recently extended to temporal settings with momentary CI tests, introduced as PCMCI~\cite{runge2019detecting}, which proved particularly robust to autocorrelated data, but limited to detecting lagged relationships. This limitation has been addressed with PCMCI+ that also handles instantaneous connections~\cite{runge2020discovering}.

\paragraph{Functional-based}
The lack of causal graph identifiability from observational data when using constraint-based methods has led to new approaches with stronger assumptions pertaining to the functional form of the data generating process, leading to identifiability guarantees. A foundational work in this category, \cite{shimizu2006linear}, introduced identifiable models based on linearity and non-Gaussianity (LiNGAM). This core idea was later extended to time series with VARLiNGAM~\cite{hyvarinen2010estimation}. Another notable work in functional models instead assumes additive noise models to obtain identifiability~\cite{hoyer2008nonlinear}, which was later also applied to temporal problems as TiMINo~\cite{peters2013causal}.

\paragraph{Gradient-based}
The challenging nature of learning causal graphs from data (i.e., discrete space) has been addressed with NOTEARS~\cite{zheng2018dags} by casting the (acyclic) graph learning task into the continuous space, notably enabling the use of conventional optimizers, such as stochastic gradient descent, in the graph inference task. This idea has been further extended to time series with DYNOTEARS~\cite{pamfil2020dynotears} that simultaneously learns instantaneous and lagged relationships.

\paragraph{Cross Mapping} Methods of this category assume that the observed time series were generated by a latent deterministic dynamical systems, which is a class of problems that GC fails on \cite{sugihara2012detecting}. Various versions of Takens' theorem assert that the latent state-space can be reconstructed from a bank of time-lagged filters under mild assumptions \cite{sauer1991embedology}. A cross mapping between these reconstructions is then interpreted as a signature of causal influence. Convergent cross mapping (CCM), \cite{sugihara2012detecting}, is the most popular CCD method based on this logic. Like GC, other methods of this type use different test statistics to infer the existence of a cross map \cite{benkHo2024bayesian,butler2024tangent}. Although the restrictive assumptions required for cross mapping methods may hinder their generality \cite{butler2023causal}, cross mapping methods are good examples of CCD methods that are neither based on the GC principle nor causal ML algorithms.

The methods GC, TE, and CCM originated from fields such as econometrics, neuroscience and climate science, with a specific focus on analysing time series, and are somewhat established within the signal processing literature. In contrast, methods such as PCMCI, DYNOTEARS, and VARLiNGAM emerged from the causal ML literature, generalizing approaches for causal discovery on static data to time series. Although developed separately, signal processing and causal ML approaches to the CCD task are not unrelated. On the contrary, we argue that all of these approaches are related, and require similar considerations for hyperparameter selection.

\section{Problem Formulation}
We now introduce some basic notation, and then introduce the problem of temporal robustness of CCD algorithms. 

\subsection{Notation}
Let $\bx_t = [x_{1,t}, \ldots, x_{D,t}]^\top \in \mathbb{R}^D$ denote a multivariate signal observed at discrete time $t$. To frame the problem, we consider models of the following form. The CCD methods described in Sec. \ref{sec:methods} do not necessarily use these models in their formulation, but it is helpful to specify a model to clarify the scope of our task. Let us assume a data generative process,
\begin{equation}
  \bx_t = \bF(\bx_{t-1}, \ldots, \bx_{t-Q},\bw_t),
  \label{eq:dgp}
\end{equation}
where $\bF$ is a vector-valued function depending on previous values of the multivariate signal $\bx_t$ and an exogeneous noise variable, $\bw_t \sim p(\bw_t)$.
The parameter $Q$ is known as the \textbf{window length}, which measures the amount of memory (measured in discrete time) that the model has. 
 
We imagine the discrete times to correspond to regular observations of a real-world physical process that is observed with sampling rate, $f_S$. In some applications, changing $f_S$ is impossible because only data is received and the modeller has no access to the recording instrument. However, even in such cases, data may be record in a high-resolution format and then downsampled as a part of preprocessing. As such, we will also discuss the effect of modifying the sampling rate as a parameter. Modification of the sampling is often done by downsampling, where we define the \textbf{downsampling factor}, 
$$
k = f_S^{original}/f_S^{new}.
$$
Finally, it is sometimes more convenient to discuss sampling rate in terms of its reciprocal, the \textbf{sampling period}, denoted
$$
\tau = 1/f_S.
$$

\begin{figure}[t]
    \centering
    \includegraphics[width=0.95\linewidth]{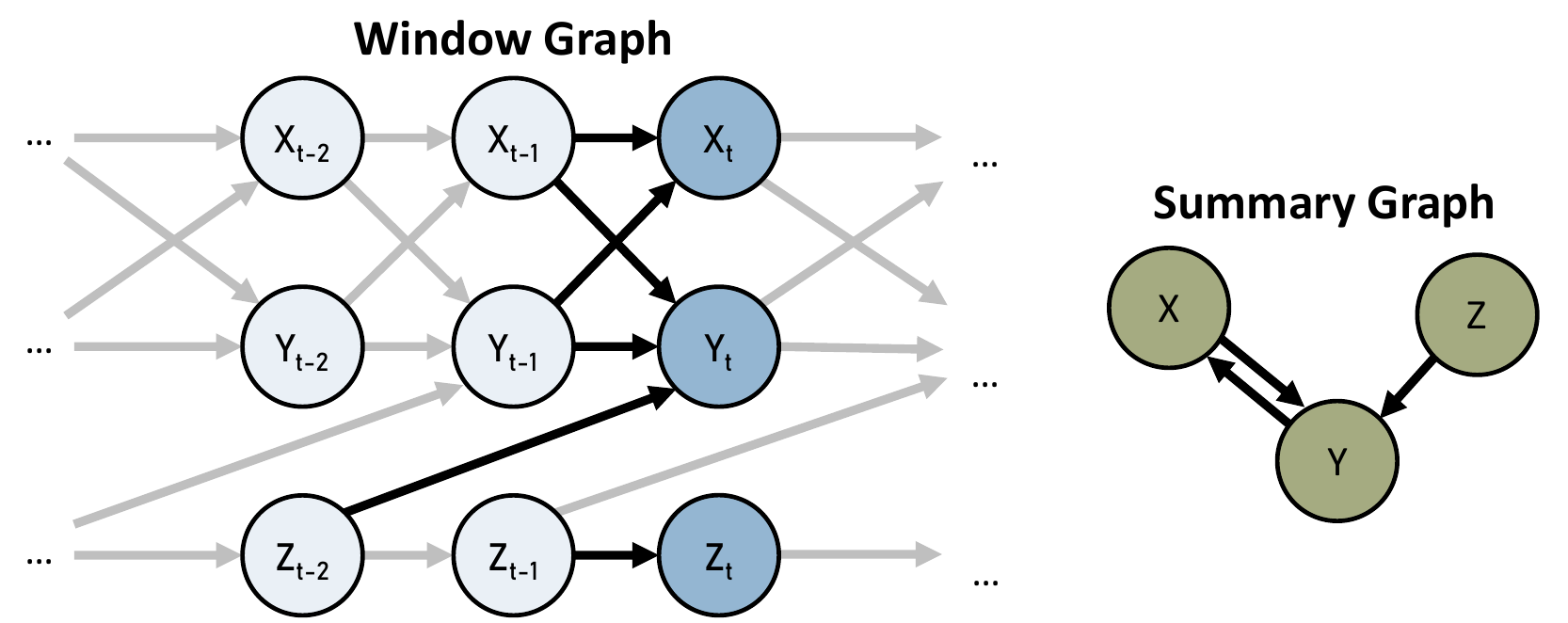}
    \caption{Descriptions of a causal system using a window graph and a summary graph, respectively. A summary graph could be induced by a window graph by aggregating over the edges in a window graph.}
    \label{fig:graphs}
\end{figure}

\subsection{Chronological Causal Discovery}
In the CCD task, we attempt to estimate the topology of a graph that is faithful to the data-generative process. Several methods to perform the CCD task are mentioned in Sec. \ref{sec:methods}. These methods can be separately into two categories depending on the type of graph that they output, shown in Fig. \ref{fig:graphs}. \textbf{Window graphs} are directed graphs over a set of nodes $x_{i,t}, i=1,...,D$ and their time lags $x_{i,t-q}$ for $q=1,...,Q$. In comparison, a \textbf{summary graph} only has $D$ nodes, corresponding to each time series $x_{i,t}$.  Edges in the window graph specify the time delay associated with a causal relationship, while the summary graph describes that there exists some causal relationship, involving one or more time lags. For methods such as PCMCI, DYNOTEARS and VARLiNGAM, a window graph is used to describe causal relationships. Meanwhile, the outputs of GC, TE, and CCM are better described as summary graphs.

\subsection{Temporal Robustness of CCD Methods}
One challenge of causal reasoning with statistical models is that there is an important distinction between the true data generative process (DGP) and our model of that process. In other tasks, like prediction, classification, or generative modelling, this distinction is less important since those tasks only require that the model match the statistics of the DGP, regardless of how the model is implemented. In causal modelling, however, we have less freedom in that we need the statistical model to more accurately mimic the DGP. 

\paragraph{A case study from neuroscience}
In real systems, any flow of information comes with some time delay, as even light is not capable of instantaneous transportation. As such, if one wishes to detect a causal influence between two processes, then the modeller must cast the right net. For example, in \cite{butler2023approach}, the task was to detect causal influence between two brain regions. As the window length $Q$ was varied, the strength of causal influence, as estimated by the model, also varied. The true time delay associated with the causal influence was hypothesized to be around 8 milliseconds, according to expert knowledge. Because the sampling rate was $\tau=1$ millisecond, it is then expected that a model with window length near $Q\approx 8$ would be appropriate to detect the effect. It was observed that for small values of $Q$, only a weak causal influence could be detected. Meanwhile, for $Q \gg 8$, the estimated causal strength was also weaker. For values near $Q=8$, there was a stronger casual strength observed. The takeaway from this study, for the current discussion, is that there is a limited range of values of $Q$ for which causality can be detected. 

\paragraph{Effects of sampling rate}
Crucially for the model \eqref{eq:dgp} to be considered \textit{causal}, we enforce $\bF$ to only depend on past values of $\bx_t$, and not future values. If we allow $\bF$ to also take $\bx_t$ as an input, we say that the model admits \textit{instantaneous effects} \cite[p.198]{peters2017elements}. Modelling instantaneous effects requires us to impose acyclicity assumptions on graphs produced by the model. However, instantaneous effects can often be considered as artifacts of a low sampling rate, $f_S$ (or a high sampling period $\tau$).
Besides instantaneous effects, it has been found that subsampling in excess ($\downarrow f_S$) makes the performance of causal methods deteriorate \cite{munoz2023synthetic, danks2013learning, barnett2017detectability, pamfil2020dynotears}.

\paragraph{Parameter selection}
Beyond the concerns of causal modelling, parameters such as $Q$ are used in other models, and therefore there are existing methods for selecting their values. There is some temptation to assume that a $Q=1$ is enough, but often times they cannot capture complex dynamics. This is somewhat obvious for (fully observed) linear systems, since their dynamics can be expressed using a single transition matrix and only one lag \cite{kalman1963mathematical}. 

In general, $Q>1$ is required to yield a good description of a system's dynamics. Preferences for how to select $Q$ are often method dependent. A na{\"i}ve approach is to select a large value for $Q$, since graphs corresponding to a smaller value of $Q$ are typically contained within the same model class. However, this approach is not justified by any principle, and extreme choices of $Q$ may not yield good results, as we shown in Sec. \ref{sec:varying_Q_experiments}. Akaike and Bayesian information criteria are popular metrics for VARX models and their relatives \cite{stoica2004model}. This  When using methods like CCM, the method of false-nearest neighbors is often employed to select $Q$ \cite{kennel1992determining}. These methods vary significantly in their assumptions and their logic, but they share a common essence. Systems which are incompletely observed have additional degrees of freedom that cannot be modelled with just one lag, and so the additional lags are required to gain information about the dynamics and state of the unobserved system.

\section{Experimental Results: Benchmarking CCD Methods}
We now examine the empirical evidence that proper selection of $\tau$ and $Q$ matter for the CCD task. We begin by benchmarking a few common CCD algorithms from the causal ML literature. We then replicate similar phenomena using the methods of Granger causality. Taking Granger causality as a reference, we then discuss phenomena that appear in extreme choices of hyperparameter selection. 

In both experiments, a bivariate system with two variables $x$ and $y$ is simulated. The signal $x$ is generated as a smoothly varying stochastic process, and $y$ is generated using a delay filter applied to $x$, with another autoregressive noise process added to the output. If $x$ is considered to be signal, then the signal-to-noise ratio in $y$ is 80-to-20. The linear filter is designed such that the time delay from input-to-output is about 50 samples in the base sampling rate.

\subsection{GC - Varying $Q$ and $k$}
\label{sec:varying_Q_experiments}


\begin{figure}[t]
    \centering
    \includegraphics[width=0.8\linewidth]{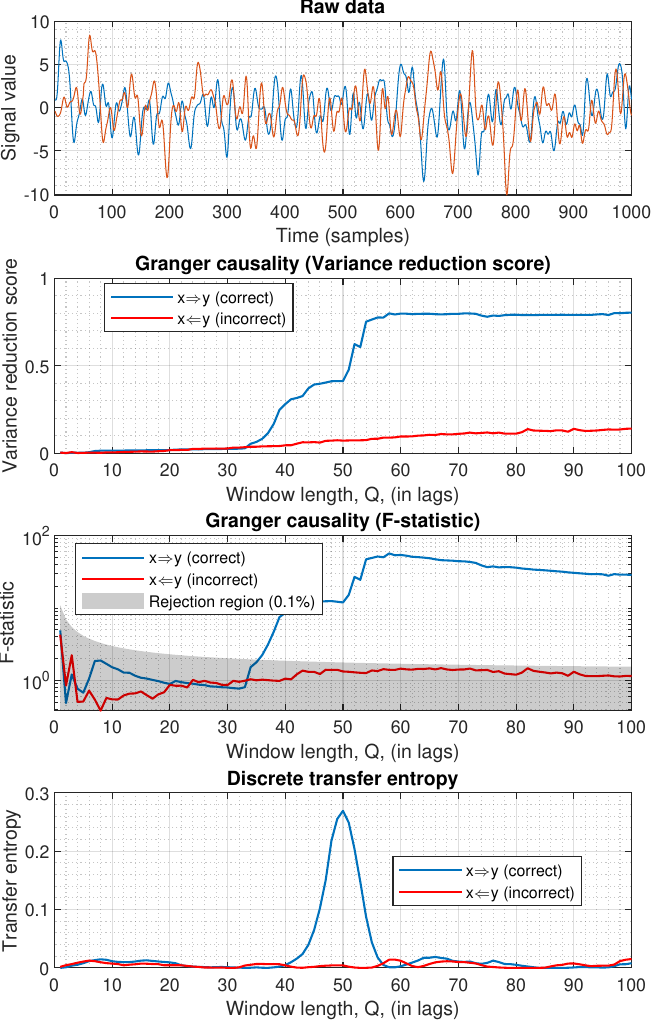}
    \caption{Varying the window length Q.}
    \label{fig:varyQ}
\end{figure}

We now take a closer look at GC and its robustness to varying the hyperparameters.
In Fig. \ref{fig:varyQ}, we visualize the performance of three different causality detectors vs the window length $Q$. We employ two different GC tests, as well as a TE version. The variance-reduction version of GC detects causation $x \Rightarrow y$ if 
$$
1- \frac{\text{Var}( y_t | x_{t-1},..., x_{t-Q}, y_{t-1},..., y_{t-Q} )}{
\text{Var}( y_t | y_{t-1},..., y_{t-Q} )} > \theta_\text{threshold},
$$
where variance is measured using the fit of a linear model. The decision threshold is selection by the user. Similarly, we also use an $F$-statistic version of the GC test that exploits that a ratio of variance estimators obeys an $F$-distribution. We visualize the decision threshold for a 0.1 percent confidence. Finally, the TE estimator uses the Shannon entropy computed from the a binary binning approach. Similar results where observed for other numbers of bins, e.g. 5 bins. The approaches based on GC show that causality cannot reliably be detected when $Q<50$. TE was only able to detect causation when $Q$ was matched to the propagation delay of the underlying causal signal. 
These findings are consistent with \cite{barnett2017detectability}.


\begin{figure}[b]
    \centering
    \includegraphics[width=0.8\linewidth]{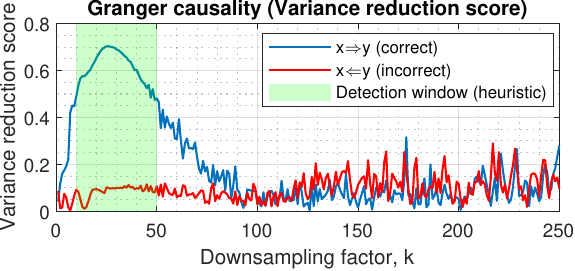}
    \caption{Varying the downsampling factor, $k$}
    \label{fig:varyK}
\end{figure}

Similarly, in Fig. \ref{fig:varyK}, we vary the sampling rate by downsampling. We employ the variance-reduction GC detector with a window length $Q=5$. In green, we visualize an approximate detection window, which is the region in which the 50-lag propagation delay is contained between $q=1$ and $q=5$ lags after downsampling by $k$. As one might hope, this heuristic anticipates where the GC test has the best results. However, the detection window cannot be computed without knowledge of the propagation delay of causal influence in the true system, which is information that is often not usually available when performing tasks like CCD.

Consistent with our previous results in Sec.~\ref{sec:ml_ccd}, GC also shows sensitivity to both hyperparameters that can determine its success or failure in detecting causality. In general, our results are also largely consistent with observations in other studies (c.f., \cite[Appendix B]{butler2024tangent} and  \cite[Appendix C]{sanchez2025causal}).

\subsection{Benchmarking ML CCD}\label{sec:ml_ccd}
\begin{figure}[t]
    \centering
    \includegraphics[width=0.95\linewidth]{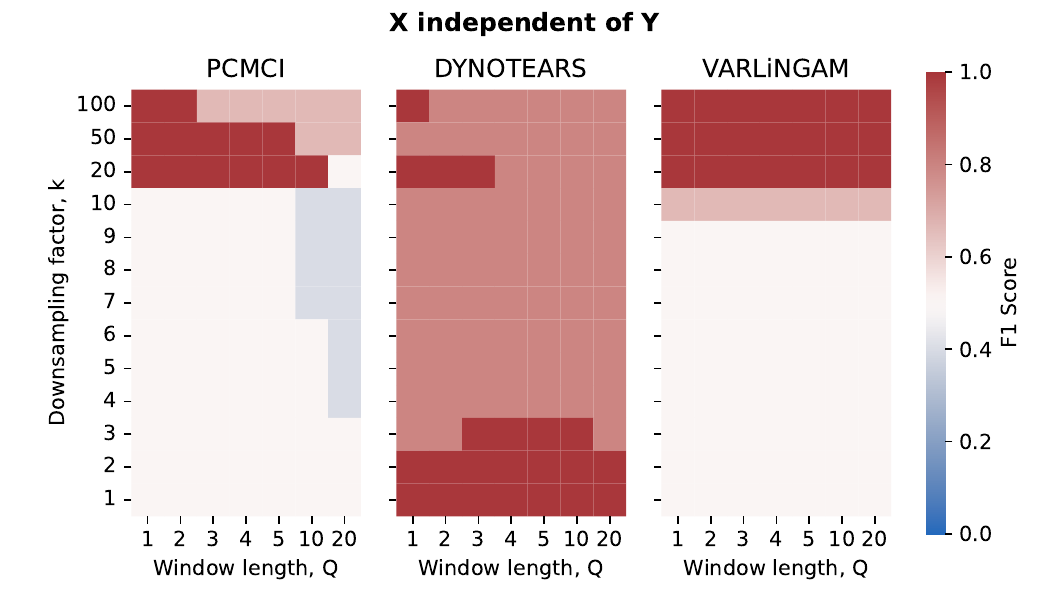}
    \caption{F1 score when X is independent of Y while varying Q and k.}
    \label{fig:indep_f1}
\end{figure}

\begin{figure}[t]
    \centering
    \includegraphics[width=0.95\linewidth]{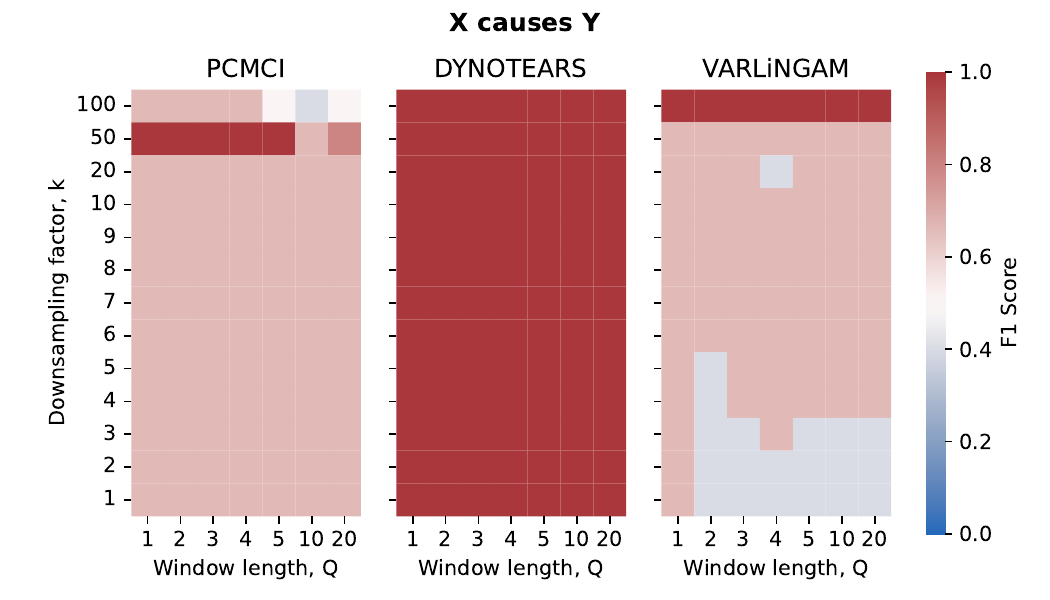}
    \caption{F1 score when X causes Y while varying Q and k.}
    \label{fig:xy_f1}
\end{figure}

We measure the effectiveness of inferring the correct causal relationship between X and Y with F1 score. Below we define F1 using (R)ecall and (P)recision, which in turn take into account true positives (TPs), false negatives (FNs) and false positives (FPs) obtained from comparing the true and predicted adjacency matrix encoding the causal graph.
\begin{align*}
    \text{R} = \frac{TP}{TP + FN}, && \text{P} = \frac{TP}{TP + FP}, && \text{F1} = \frac{2 \times \text{R} \times \text{P}}{\text{R} + \text{P}}.
\end{align*}
Intuitively, higher F1 is better and $\text{F1}{=}1$ indicates perfectly recovered graph. We consider two settings wherein X either is independent of, or causes, Y, results of which are depicted in Fig. \ref{fig:indep_f1} and \ref{fig:xy_f1} respectively.

From both cases it is clear that varying hyperparameters Q and k has a large impact on the inferred causalities (i.e., colors change across Q and k). The right values for Q and k that lead to the right graph (dark red) also depend on used CCD algorithm, which is consistent with a similar observation from causal discovery on static data that hyperparameter robustness can be algorithm-specific~\cite{machlanski2024robustness}.

\section{Discussion}
Discussions of sampling rate selection often consider the Nyquist-Shannon theorem to produce lower-bounds on sampling rates that allow for lossless signal reconstruction \cite{zeng2024generalized}. In a similar spirit, a \textit{causal Nyquist theorem} would consider the spectral properties of a causal mechanism, and provide a lower-bound on the sampling rate required for correct causal inference. 
One approach, assuming we model causation via linear time-invariant filters, is to study transfer functions \cite{Jalili_Candelino_2023}.
Nyquist theorems typically assume that frequency spectra are band-limited. If the transfer function of a linear filter is band-limited, then one could define a Nyquist threshold for sampling rates, which would upper-bound the Nyquist rate of any signal output by the filter. However, not all linear filters are band-limited; notably, high-pass filters have the complete opposite behaviour. In such cases, it is unclear how one might define a Nyquist rate.
Furthermore, although sub-Nyquist sampling can introduce artifacts, it may be possible that CCD performance is not adversely affected. It remains to be seen if a good notion of a causal Nyquist rate can be defined.

\section{Conclusion}
Our observations point to challenges that CCD methods face when applied to real data. These challenges apply to both traditional and modern methods alike. These challenges were also observed despite our limited problem scope, ignoring complexities that arise from nonstationarity, latent confounder processes, and multiple time scales. The signal processing community has clear opportunities to contribute to the overcoming these challenges and developing better CCD methods. 

\section{Acknowledgement}
We thank several individuals for useful discussions: Emily Stephen, Urbashi Mitra, Michael Camilleri, and Petar Djuri\'{c}.

\bibliographystyle{ieeetr}
\bibliography{refs}

\end{document}